\documentclass[runningheads]{llncs}
\usepackage{graphicx}

\usepackage{tabularx}
\newcolumntype{C}[1]{>{\centering\arraybackslash}p{#1}}
\usepackage{multirow}
\usepackage{hhline}
\usepackage{adjustbox}
\begin{document}
\title{Lung Cancer Tumor Region Segmentation Using Recurrent 3D-DenseUNet\thanks{source code available at https://github.com/muntakimrafi/TIA2020-Recurrent-3D-DenseUNet}}
\titlerunning{Recurrent 3D-DenseUNet}
% If the paper title is too long for the running head, you can set
% an abbreviated paper title here
%
\author{Uday Kamal\inst{1} \and
Abdul Muntakim Rafi\inst{2} \and Rakibul Hoque\inst{1} \and Jonathan Wu\inst{2} \and
Md. Kamrul Hasan\inst{1}}
\authorrunning{Kamal et al.}
% First names are abbreviated in the running head.
% If there are more than two authors, 'et al.' is used.
%
\institute{Bangladesh University of Engineering and Technology, Dhaka-1205, Bangladesh \and
University of Windsor, 401 Sunset Ave, Windsor, ON N9B 3P4, Canada\\
\email{udday2014@gmail.com, \{rafi11, jwu\}@uwindsor.ca, khasan@eee.buet.ac.bd}}
\maketitle              % typeset the header of the contribution
\begin{abstract}

The performance of a computer-aided automated diagnosis system of lung cancer from Computed Tomography (CT) volumetric images greatly depends on the accurate detection and segmentation of tumor regions. In this paper, we present Recurrent 3D-DenseUNet, a novel deep learning based architecture for volumetric lung tumor segmentation from CT scans. The proposed architecture consists of a 3D encoder block that learns to extract fine-grained spatial and coarse-grained temporal features, a recurrent block of multiple Convolutional Long Short-Term Memory (ConvLSTM) layers to extract fine-grained spatio-temporal information, and finally a 3D decoder block to reconstruct the desired volume segmentation masks from the latent feature space. The encoder and decoder blocks consist of several 3D-convolutional layers that are densely connected among themselves so that necessary feature aggregation can occur throughout the network. During prediction, we apply selective thresholding followed by morphological operation, on top of the network prediction, to better differentiate between tumorous and non-tumorous image-slices, which shows more promise than only thresholding-based approaches. We train and test our network on the NSCLC-Radiomics dataset of 300 patients, provided by The Cancer Imaging Archive (TCIA) for the 2018 IEEE VIP Cup. Moreover, we perform an extensive ablation study of different loss functions in practice for this task. The proposed network outperforms other state-of-the-art 3D segmentation architectures with an average dice score of 0.7228.

\end{abstract}
\section{\textbf{Introduction}}
Lung cancer, also known as \textit{lung carcinoma}, is a malignant lung tumor characterized by uncontrolled cell growth in tissues of the lung. The abnormal cells do not develop into healthy lung tissue. Instead, they divide rapidly and form tumors. Researchers have found that it takes a series of DNA mutations to create a lung cancer cell. It can be caused by the normal aging process or environmental factors, such as cigarette smoke, breathing in asbestos fibers, and exposure to radon gas \cite{pataer2012histopathologic}. There are two principal types of primary lung cancer, non-small cell lung cancer (NSCLC) and small cell lung cancer (SCLC). These two types are quite different in their behavior and response to treatments. The most common form among these is the NSCLC \cite{uzelaltinbulat2017lung}, which accounts for about 85\% of all lung cancers.

The primary diagnosis of lung cancer is performed using different non-invasive screening procedures, such as CT scan or Magnetic Resonance Imaging (MRI). A radiologist confirms the existence of malicious cancer cells or tumors from the scanned images before conducting a more formal biopsy test. He extracts and analyzes several features (e.g., attenuation, shape, size, and location) from the medical images. For this purpose, CT scan is a more affordable choice than the MRI. However, it is a daunting task to detect lung tumors from CT scans, and they can often be overlooked \cite{del2017missed}. Moreover, the developing and undeveloped countries face a scarcity of expert radiologists, and lung cancers are detected at the final stages in most cases \cite{midthun2016early}. In this regard, computer-aided diagnosis (CAD) tools have become a necessity to assist the radiologists in making decisions from the raw data of a CT scan quickly with better accuracy.

Several methods have been proposed in the literature to complete the task of lung tumor segmentation. Uzelaltinbulata and Ugur \cite{uzelaltinbulat2017lung} have proposed an image processing based technique to segment the lung tumor in CT images. They have used morphological operations, filtering, seeding, thresholding, and image residue to develop a system that automatically segments lung tumors. Aerts et al. \cite{aerts2014decoding} present a radiomic analysis of 440 features that quantify tumor image intensity, shape, and texture, extracted from CT data. In \cite{lambin2012radiomics}, the authors have discussed different statistical approaches to analyze CT data and the associated challenges. Besides lung tumor, a lung image contains many other components, or structures, such as clavicles and lungs. Therefore, detection of a lung tumor can be easily misguided by techniques that solely depend on the image processing technique, where pixel intensity and thresholding are heavily utilized to differentiate the tumor region. Towards this end, researchers are now more focused on machine learning based methods such as Deep Neural Networks(DNN) to solve medical image segmentation problems. These data-driven approaches have radically improved the performance of tumor detection compared to the complete image processing based techniques. 

In \cite{milletari2016v}, the authors have introduced V-Net, a fully convolutional neural network (CNN), as well as a unique objective function to conduct volumetric segmentation from MRI \cite{milletari2016v}. Cicek et al. \cite{cciccek20163d} perform volumetric segmentation through a successful adaptation of U-Net \cite{ronneberger2015u} by replacing all the 2D operations with their 3D counterparts. In \cite{chen2016voxresnet}, the authors propose a deep voxel-wise neural network `VoxResNet,' where they explore residual learning for a 3D volumetric segmentation task. Liao et al. \cite{liao2013representation} have demonstrated the application of a stacked independent subspace analysis (ISA) network to perform 3D prostate segmentation. Another interesting work is the proposition of a novel Hybrid-DenseUNet \cite{li2017h}, where the authors utilize the sophisticated connectivity of DenseNet \cite{huang2017densely} in the U-Net architecture. The dense connections convey associated input features into the deeper levels of the U-Net architecture, which may otherwise suffer from being diminished during the forward propagation. Hossain et al. \cite{hossain2019pipeline} propose 3D LungNet for performing lung tumor segmentation, where they incorporate a hybrid training method to train their network. In \cite{pang2019fast}, the authors adopt a transfer learning strategy for performing segmentation to alleviate the noisy pixel-level label interruption. Pang et al. \cite{pang2020ctumorgan} propose a unique adversarial learning framework for segmentation. Despite the breadth of works done in the field of volumetric tumor segmentation, none of these approaches have considered the end-to-end recurrent features inherent to the volume CT-scan data. Most of these works heavily rely on the 3D convolutional layers that extract coarse-grained temporal features by utilizing only the neighboring frames instead of the complete volume. 

In this paper, we propose a novel Recurrent 3D-DenseUNet architecture for volumetric lung tumor segmentation from CT scans. The core structure of our network is inspired by the widely popular U-Net architecture \cite{ronneberger2015u} that consists of a traditional encoder-decoder structure. We modify the network by replacing the 2D convolution operations with their 3D counterparts, except for the pooling layers. Instead of the conventional spatio-temporal pooling operation followed by 3D convolution, we adopt spatial-only pooling layers in order to better preserve the temporal features across slices. We also incorporate the dense connectivity proposed in \cite{huang2017densely} within our architecture to enable more fine-grained and aggregated feature propagation. In order to transform the coarse-grained output of the 3D encoder into a richer and fine-grained spatio-temporal feature space, we incorporate a recurrent block consisting of several convolutional long short-term memory (ConvLSTM) layers \cite{hochreiter1997long} in between the encoder-decoder structure. In addition, the encoder-decoder blocks are interconnected by a hierarchical set of skip connections so that any necessary features lost through the encoding operation can be regenerated during the reconstruction procedure performed by the decoder. The final binary segmentation mask is generated by thresholding the model output followed by dilation, a morphological operation, to reduce the pixel-level anomaly present in the prediction for better representation of the tumor region.

The rest of the paper is organized as follows. Section 2 provides a detailed description of our proposed network. We offer a detailed description of our training procedures, evaluation criteria, experimental results, and ablation study in Section 3. We conclude in section 4.

\section{\textbf{Proposed Network Architecture}}

\begin{figure*}
	\includegraphics[width=4.5in]{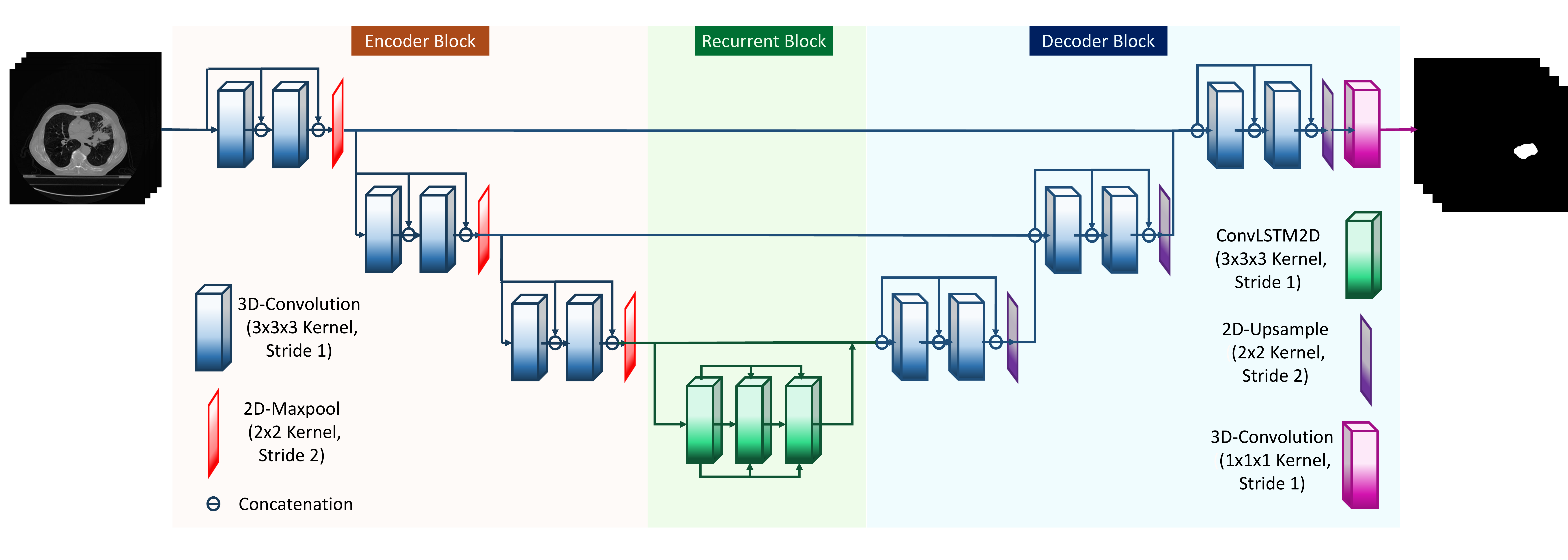}
	\caption{The architecture of our proposed Recurrent 3D-DenseUNet model.}
	\label{rakib3}
\end{figure*}

In this work, we propose a novel architecture that harnesses the core essence of three fundamental networks: DenseNet \cite{huang2017densely}, U-Net \cite{ronneberger2015u}, and Convolutional Recurrent Network. Our proposed model is illustrated in Fig. \ref{rakib3} and the outline of the architecture is presented in Table \ref{UConvnet}.

\begin{table}[!b]
\centering

\caption{\textbf{The outline of our proposed Recurrent 3D-DenseUNet model}}
\label{UConvnet}
\begin{adjustbox}{max width=1\textwidth,center}
\begin{tabular}{|l|l|l|l|l|l|}
\hline
\textbf{Block} &
\textbf{Sub-block/Layer} & 
\textbf{Output size} & 
\textbf{Details (Layer and Activation)} & 
\textbf{Kernels} \\
\hline 
\multirow{6}{*}{{\begin{tabular}[l]{@{}l@{}}Encoder\\block \end{tabular}}} & 
{\begin{tabular}[l]{@{}l@{}}convolutional block \end{tabular}} & 256$\times$256$\times$8$\times$32  &  [Conv3D, BN, ReLU, SD] $\times2$ & F=32, k=3$\times$3$\times$3, S=1 \\ \cline{2-5}

& Maxpool2D & 128$\times$128$\times$8$\times$32 & --- & k=2$\times$2, S=2\\ \cline{2-5}

& convolutional block  & 128$\times$128$\times$8$\times$64  & [Conv3D, BN, ReLU, SD] $\times2$ & F=64, k=3$\times$3$\times$3, S=1 \\ \cline{2-5}

& Maxpool2D & 
64$\times$64$\times$8$\times$64 & --- & k=2$\times$2, S=2 \\ \cline{2-5}

& {\begin{tabular}[c]{@{}c@{}}convolutional block \end{tabular}} & 64$\times$64$\times$8$\times$128  & [Conv3D, BN, ReLU, SD] $\times2$ & F=128, k=3$\times$3$\times$3, S=1
\\ \cline{2-5}

& {\begin{tabular}[c]{@{}c@{}}Maxpool2D \end{tabular}}
& 32$\times$32$\times$8$\times$128 & --- & k=2$\times$2, S=2
\\ \hline

% recurrent block starts here
\multirow{3}{*}{{\begin{tabular}[l]{@{}l@{}}Recurrent\\block \end{tabular}}} & 
ConvLSTM2D & 32$\times$32$\times$8$\times$256 & --- & F=256, k=3$\times$3, S=1 
\\ \cline{2-5}
& {\begin{tabular}[c]{@{}c@{}}ConvLSTM2D \end{tabular}}
& 32$\times$32$\times$8$\times$256 & --- & F=256, k=3$\times$3, S=1 
\\ \cline{2-5}
& {\begin{tabular}[c]{@{}c@{}}ConvLSTM2D \end{tabular}}
& 32$\times$32$\times$8$\times$256 & --- & F=256, k=3$\times$3, S=1
\\ \hline

% decoder starts here
\multirow{7}{*}{{\begin{tabular}[l]{@{}l@{}}Decoder\\block \end{tabular}}} &  
convolutional block  & 32$\times$32$\times$8$\times$128  & [Conv3D, BN, ReLU] $\times2$ & F=128, k=3$\times$3, S=1 \\ \cline{2-5}

& {\begin{tabular}[c]{@{}c@{}}Upsampling2D\end{tabular}}
& 64$\times$64$\times$8$\times$128 & --- & {\begin{tabular}[c]{@{}c@{}}k=2$\times$2, S=2
\end{tabular}}
\\  \cline{2-5}

& {\begin{tabular}[c]{@{}c@{}}convolutional block \end{tabular}} & 64$\times$64$\times$8$\times$64  & [Conv3D, BN, ReLU] $\times2$ & F=64, k=3$\times$3, S=1 \\  \cline{2-5}

& Upsampling2D & 128$\times$128$\times$8$\times$64 & --- & k=2$\times$2, S=2
\\  \cline{2-5}

& convolutional block & 128$\times$128$\times$8$\times$32  & [Conv3D, BN, ReLU]$\times2$ & F=32, k=3$\times$3, S=1 \\  \cline{2-5}

& {\begin{tabular}[c]{@{}c@{}}Upsampling2D\end{tabular}}
& 256$\times$256$\times$8$\times$32 & --- & {\begin{tabular}[c]{@{}c@{}}k=2$\times$2, S=2
\end{tabular}}
\\  \cline{2-5}
& Conv3D & 256$\times$256$\times$8$\times$1 & Sigmoid & F=1, k=1$\times$1, S=1

\\ \hline

\end{tabular}
	\centering 
\end{adjustbox}
\end{table}

U-Net \cite{ronneberger2015u} is a very popular and well-established network for medical image segmentation and classification problems. Therefore, it is an obvious choice for us to experiment with U-Net for lung tumor segmentation. Thus, in this architecture, we have adopted the U-Net as the core skeleton of our network. Overall, our network consists of three major parts: Encoder, Recurrent block, and a Decoder. The details of each block are as follows:

\subsubsection{\textbf{Encoder Block}}
The encoder of our network comes with several convolutional blocks, each consisting of two subsequent 3D-Convolutional layers with kernel size $(3\times3\times3)$, a Batch-Normalization layer, ReLU as an Activation layer and 2D-Maxpooling operation with a kernel size of $(2\times2)$. The main reason for not using a 3D-pooling operation is to preserve temporal information as much as possible. A spatial dropout layer with a dropout rate of 10\% has also been used at the end of every convolutional block in order to prevent overfitting on the training dataset. Since we are interested in pixel-level segmentation, the feature space we are more interested in needs to contain as much high-level information as possible. However, degradation of features strength due to forward propagation of the input data is unavoidable in any feedforward neural network unless the skip connection technique is applied. As shown in \cite{huang2017densely}, dense-blocks, which are the building blocks of DenseNet, are interconnected with each other in the network and thereby prevent loss of acute statistical features due to forward propagation. Inspired by this idea, each convolutional block in the encoder is designed to perform as pseudo-dense-block, which establish inter-connectivity between the input and middle layers of the block. 

\subsubsection{\textbf{Recurrent Block}}
The main intuition of using volume data is to capture the inter-slice continuity of the tumor as a solid object. For this purpose, we have used a recurrent block consisting of several ConvLSTM layers \cite{xingjian2015convolutional} within the transition section from the encoder to the decoder. With the increase in depth of convolutional layers, higher and higher level features are extracted from the input. To capture the interdependencies at the highest level of our network, we have used the recurrent block at the transition section of encoder and decoder, which helps the network to capture more fine-grained spatio-temporal features and inter-dependencies among the output features of the encoder. Each ConvLSTM layer is comprised of 2D-convolutional layers with kernel size $(3\times3\times3)$.

\subsubsection{\textbf{Decoder Block}}
The decoder block is the final part of the network that takes the high-level spatio-temporal features captured by the recurrent block as input and generates the volumetric segmentation mask as output. The decoder has a similar architectural structure of the encoder, except the spatial pooling layer is replaced by a spatial Upsample layer that performs the brings the latent feature space to the original input resolution. At decoder block output, a skip connect is established with the corresponding previous encoder block. Finally, a pointwise 3D convolution layer with Sigmoid activation is used to bring the volume mask back to its original input dimension.

\section{\textbf{Experimental Analysis}}

We perform several experiments to demonstrate the efficacy of our proposed method. In this section, we discuss those experimental procedures and results in detail. All of the experiments regarding training and implementation of the model are performed in hardware environments, which included Intel Core-i7 8700K, 3.70 GHz CPU, and Nvidia GeForce GTX 1080 Ti (11 GB Memory) GPU. The necessary codes are written in Python and, the neural network models are implemented by using the Keras API with Tensorflow-GPU in the backend.

\subsection{Description of the Dataset}

The dataset we use in our paper consists of 300 patients from the NSCLC-Radiomics dataset \cite{clark2013cancer}. The contour-level annotations for these 300 patients have been prepared for the IEEE VIP Cup 2018 with the help of an expert radiologist \cite{mohammadi2018lung}. The dataset contains scanned 3D volumes of the chest region for each patient. It is provided in DICOM format with slices of $512 \times 512$ size. We divide the dataset into two sets- train (260) and test (40), the same way as \cite{hossain2019pipeline}. We use 10\% of the training data as our validation set. The distribution of the tumorous and non-tumorous slices are provided in Table \ref{tab_dist}.

\begin{table}[!t]
	\centering
	\caption{Number of different types of CT scan slices in our dataset}
	\begin{tabular}{|C{1.1in} | C{1.1in} | C{1.1in} | C{1.1in}|}
		\hline
		Data & Number of Patients & Tumor & Without Tumor \\ 
		\hline
		Train & 260 & 4296 & 26951 \\ 
		\hline
		Test & 40 & 848 & 3630 \\ 
		\hline
	\end{tabular}
	
	\label{tab_dist}
\end{table}

\begin{figure}
    \centering
	\includegraphics[width=4.5in]{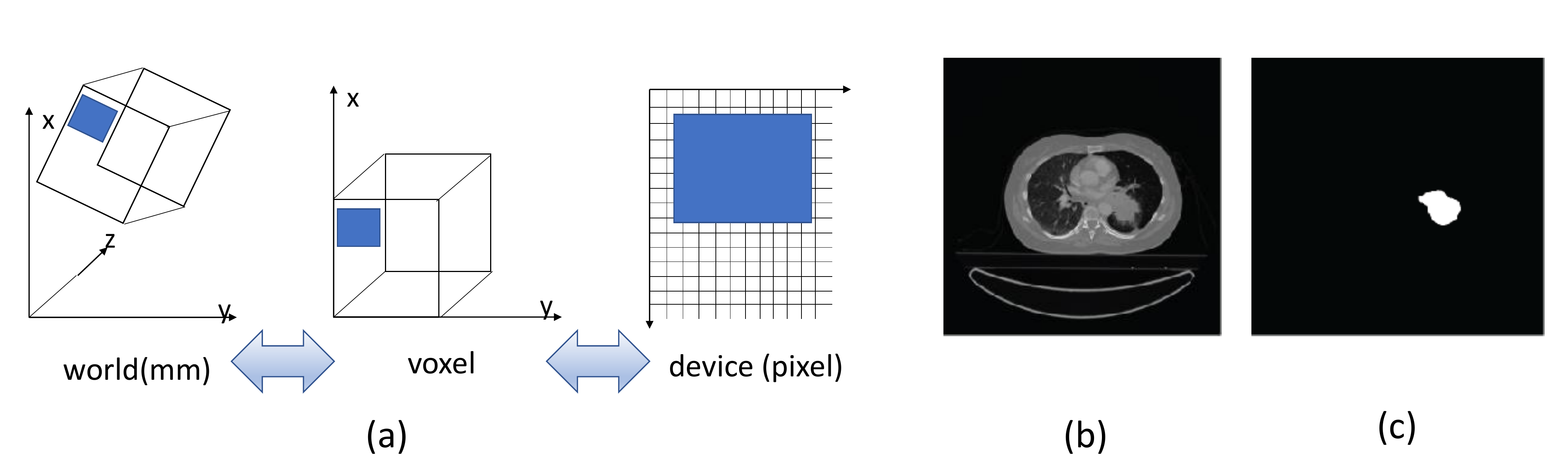}
	\caption{Data generation steps from a DICOM file. \textbf{(a)} Conversion of DICOM files into pixel images. \textbf{(b)} A sample 2D-grayscale image of size $512\times512$ extracted from the DICOM files. \textbf{(c)} A binary mask of tumor for the image presented in \emph{(b)}.}
	\label{lung1}
\end{figure}

We convert the DICOM data to 2D gray-scale images using the Pydicom \cite{mason2011t} Library and extract their respective binary masks from corresponding RTconstruct files. We resize the images from ($512\times512$) to ($256\times256$) to cope with the memory limitation of GPU. Afterward, we create patches of \emph{eight} by concatenating eight consecutive images. We can observe in Table \ref{tab_dist} that the dataset suffers from severe class imbalance. A large percentage of the CT scan slices do not contain any tumorous region. Moreover, the larger portion of slices containing tumors has non-tumorous regions, as shown in Fig. \ref{lung1}. Therefore, during training, we only use the patches $(256\times256\times8)$ that contain at least one slice with a tumor. Here, the number of slices in a patch is also limited by the computational memory of the GPU. Also, it reduces the training parameters of the network. 

\subsection{Training procedure}

We have used several augmentation schemes to show different data to our neural network in each epoch during training. It helps to prevent the network from overfitting on the training data \cite{shorten2019survey}. The augmentations we use are as follows: random rotation, random cropping, random global shifting, random global scaling, random noise addition, random noise multiplication, horizontal flip, and blurring. They are applied online randomly during training.

We have trained our Recurrent 3D-DenseUNet model using an initial learning rate of $10^{-4}$ and batch size of \emph{2}. We do not use a greater batch size number as the computational memory of the GPU limits it. The weights of the different layers are initialized randomly with the uniform distribution proposed by Glorot and Bengio \cite{glorot2010understanding}. We use binary cross-entropy as the loss function and Adam \cite{kingma2014adam} as the optimizer with the exponential decay rate factors $\beta_1$ = 0.9 and $\beta_2$ = 0.999. We decrease the learning rate by a factor of 0.5 if the validation dice coefficient does not increase in three successive epochs. We train our network for a maximum number of 30 epochs and select the model with the highest validation dice coefficient.

\subsection{\textbf{Evaluation Criteria}}
We use dice coefficient as our evaluation criterion to compare between generated mask and the ground truth for all test images. Dice coefficient is a measure of relative overlap, where $1$ represents perfect agreement and $0$ represents no overlap.
\begin{center} 
	\begin{equation}
	\label{eqn:eq1}
	D = \frac{2(|X| \cap |Y|)}{|X|+|Y|}
	\end{equation}
\end{center}

In (\ref{eqn:eq1}), $\cap$ denotes the intersection operator, $X$ and $Y$ are the predicted binary mask and ground truth. It should be noted that the dice coefficient $(D)$ has a restricted range of $[0,1]$. The following two conventions are considered in the computation of Dice coefficient: (i) For True-Negative (i.e., there is no tumor, and the processing algorithm correctly detected the absence of the tumor), the dice coefficient would be 1, and; (ii) For False-Positive (i.e., there is no tumor, but the processing algorithm mistakenly segmented the tumor), the dice coefficient would be 0.

\subsection{Results}

\begin{figure*}
    \centering
	\includegraphics[width=3in]{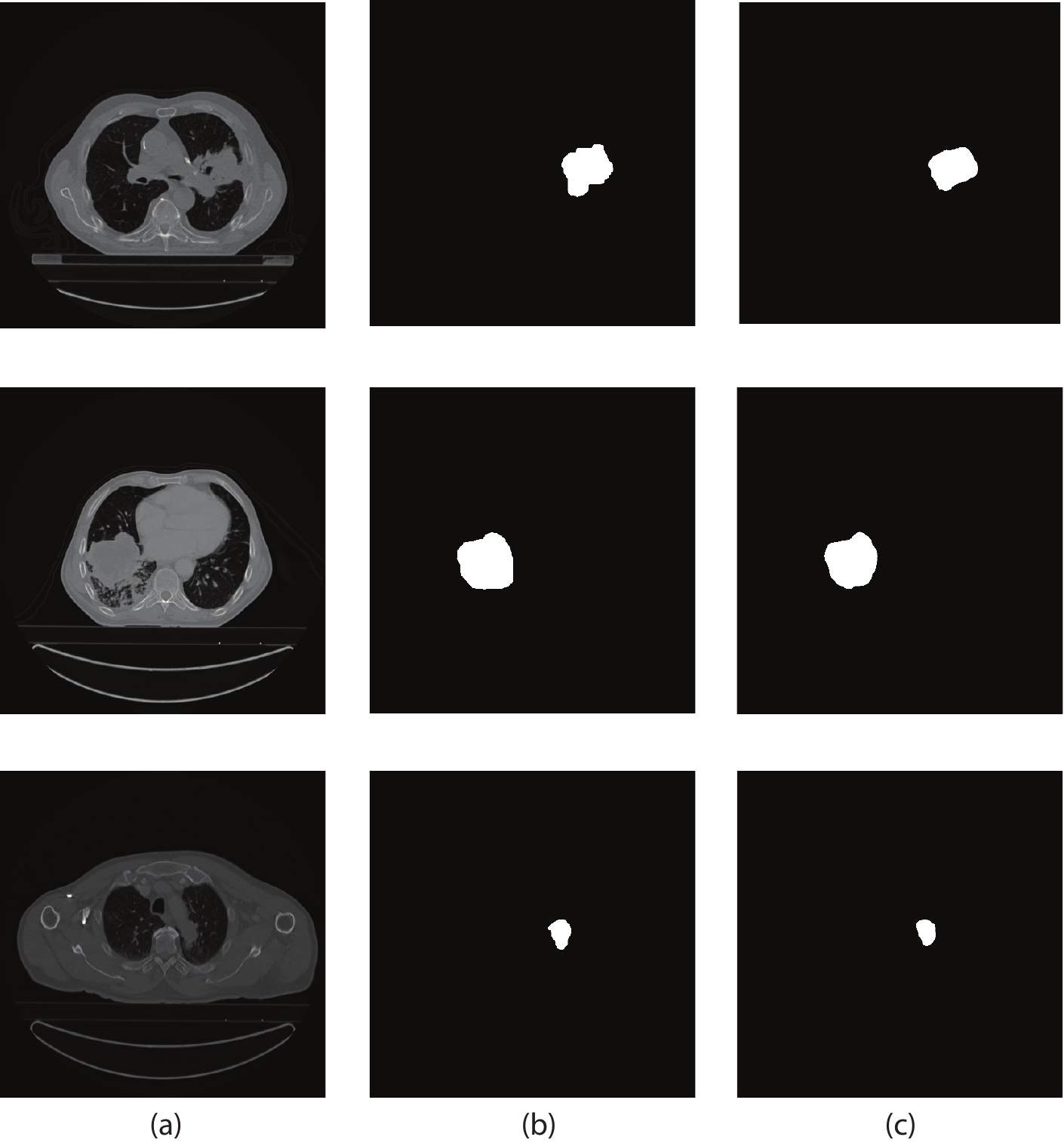}
	\caption{Comparison between ground truth and binary masks predicted by our prposed method. (a) Lung image, (b) ground truth, and (c) predicted mask.}
	\label{lung2}
\end{figure*}

During testing, we create overlapping patches of $(256 \times 256 \times 8)$ from all the CT scan slices of a single patient with stride $1$. Then we use our Recurrent 3D-DenseUNet network to generate volumetric prediction maps. For the masks that belong to multiple patches, we average over multiple predictions. The final binary segmentation mask is generated by thresholding the model output followed by dilation, a morphological operation. The hyperparameters, associated with the post-processing operations, are chosen based on the performance on the validation data. We use a 0.7 threshold to convert the network prediction to binary masks. Afterward, we apply dilation using a $(7\times7)$ circular kernel to improve the IOU (Intersection over Union) value. It reduces the pixel-level anomaly present in the prediction and provides a better representation of the tumor region. We compare our proposed method with several state-of-the-art methods available in the literature for performing segmentation tasks. The performance of the different methods is shown in Table \ref{tab_comp}. Our proposed method outperforms other approaches by a significant margin. A few predictions from the test set are shown in Fig. \ref{lung2}.

\begin{table}[!t]
	\centering
	\caption{Dice coefficient for different methods in lung tumor segmentation task}
	\begin{tabular}{|C{1.5in} | C{1.5in} | C{1.5in}|}  
		\hline
		Model & Mean Dice coefficient & Median Dice coefficient \\ 
		\hline
		2D-UNet \cite{ronneberger2015u} & 0.5848 & 0.6229 \\ 
		\hline
		2D-LungNet \cite{anthimopoulos2018semantic} & 0.62.67 & 0.6678 \\
		\hline
		3D-LungNet \cite{hossain2019pipeline} & 0.6577 & 0.7039 \\
		\hline
		Proposed Method & 0.7228 & 0.7556 \\
		\hline
	\end{tabular}
	
	\label{tab_comp}
\end{table}

\subsection{\textbf{Ablation study}} 

In this subsection, we perform an ablation study of our proposed methodology and see how different architectural,  training, and testing design choices affect our performance. First, we compare the performances of different variants of our network architecture (see Table \ref{tab_design}).

\begin{table}[!t]
	\centering
	\caption{Dice coefficient for different architectures similar to ours in lung tumor segmentation task}
	\begin{tabular}{|C{1.5in} | C{1.5in} | C{1.5in}|}  
		\hline
		Model & Mean Dice coefficient & Median Dice coefficient \\ 
		\hline
		3D-UNet & 0.6961 & 0.7326 \\ 
		\hline
		3D-DenseUNet & 0.6884 & 0.7188 \\
		\hline
		3D-DenseUNet with recurrent block & 0.7049 & 0.7646 \\
		\hline
		Our proposed network  & 0.7228 & 0.7556 \\
		\hline
	\end{tabular}
	
	\label{tab_design}
\end{table}

A suitable cost function is equally important in training neural networks. The performance is determined by a combination of both the network and the loss function. Moreover, the dice coefficient may not always be the best indicator of a segmentation network's performance. The number of false positives and false negatives also play a vital role in determining the robustness of a model. Towards this end, we train and test our network with different loss functions and explore how they affect the performance of our network (see Table \ref{tab_loss}). We see that binary crossentropy outperforms the other loss functions we compare with. Also, the tversky loss function provides an opportunity to reduce the number of false positives or false negatives by changing its parameter $\alpha$ \cite{hashemi2018tversky}. It should be noted that we enable the proposed post-processing operations for the experiments shown in Table \ref{tab_loss}.

\begin{table}[!b]
	\centering
	\caption{Performance of our proposed method for different loss functions}
	\begin{tabular}{|C{0.9in} | C{0.9in} | C{0.9in}|C{0.9in}|C{0.9in}|}  
		\hline
		Model & Mean Dice coefficient & Median Dice Coefficient & False Positives & False Negatives  \\ 
		\hline
		tversky loss ($\alpha = $0.1) & 0.6244 & 0.6637 & 786 & 221 \\ 
		\hline
		tversky loss ($\alpha = $0.3) & 0.6933 & 0.7195 & 453 & 272 \\ 
		\hline
		tversky loss ($\alpha = $0.5) [Dice loss] & 0.6963 & 0.7372 & 375 & 287 \\ 
		\hline
		tversky loss ($\alpha = $0.7) & 0.7056 & 0.7282 & 417 & 324 \\ 
		\hline
		tversky loss ($\alpha = $0.9) & 0.7125 & 0.7462 & 243 & 396 \\ 
		\hline
		Focal loss \cite{lin2017focal} & 0.7159 & 0.7493 & 498 & 349 \\ 
		\hline
		IOU loss & 0.6981 & 0.726 & 437 & 296 \\ 
		\hline
		binary crossentropy & 0.7228 & 0.7556 & 321 & 331 \\ 
		\hline
		
	\end{tabular}
	
	\label{tab_loss}
\end{table}

Lastly, we empirically validate the significance of our thresholding and morphological operation. From Table \ref{tab_thresh}, we can see how the performance deteriorates without any thresholding or morphological operations. It can also be noticed that there is a trade-off among the dice coefficient, false positives, and false negatives for different thresholds. The optimum performance can be achieved with a threshold of 0.7 and dilation post-processing operation. 

\begin{table}[!t]
	\centering
	\caption{Performance of our proposed method for different thresholds and without morphological post-processing}
	\begin{tabular}{|C{0.9in} | C{0.9in} | C{0.9in}|C{0.9in}|C{0.9in}|}  
		\hline
		Model & Mean Dice coefficient & Median Dice Coefficient & False Positives & False Negatives  \\ 
		\hline
		0.5 threshold & 0.7169 & 0.7296 & 446 & 280 \\ 
		\hline
		0.6 threshold & 0.7174 & 0.7383 & 372 & 297 \\
		\hline
		0.7 threshold & 0.7228 & 0.7556 & 321 & 331 \\
		\hline
		0.8 threshold & 0.7368 & 0.7685 & 261 & 361 \\
		\hline
		0.9 threshold & 0.7278 & 0.7763 & 208 & 403\\
		\hline
		No threshold & 0.6685 & 0.7624 & 120 & 789 \\
		\hline
		No dilation with 0.7 threshold & 0.6604 & 0.7147 & 321 & 332\\
		\hline
		No dilation with 0.8 threshold & 0.6626 & 0.6859 & 261 & 362\\
		\hline
	\end{tabular}
	
	\label{tab_thresh}
\end{table}

\section{\textbf{Conclusion}}  
In this paper, we have proposed Recurrent 3D-DenseUNet, a novel 3D Recurrent encoder-decoder based Convolutional Neural Network architecture for accurately detecting and segmenting lung tumors from volumetric CT scan data. We adopt spatial-only pooling layers in our architecture, instead of the conventional spatio-temporal pooling operation to better preserve the temporal features. We also use a recurrent block consisting of several ConvLSTM layers between the encoder-decoder structure to capture the inter-slice correlation at the high-level feature space. We have also incorporated short intra-block skip connections between the input and latter layers of every encoder and decoder block throughout the network to reduce the loss of any important feature during forward propagation. We train this model using 3D volume data of size $(256\times256\times8)$ and utilize various types of data augmentations to prevent overfitting during training. Finally, we apply a threshold on top of the network prediction to create binary masks and enhance it with dilation operation. This methodology has enabled us to achieve a better dice score in lung tumor segmentation tasks and reduce the overall number of false positives and false negatives. In our future works, we want to further modify and improve the model architecture, and explore how it performs on other medical imaging tasks. 
%
% ---- Bibliography ----
%
% BibTeX users should specify bibliography style 'splncs04'.
% References will then be sorted and formatted in the correct style.
%

\bibliography{urr}
\bibliographystyle{splncs04}
\end{document}